\documentstyle[12pt]{article}

\let\ph=\varphi  \let\PH=\Phi

\def\0{\over } \def\1{\vec }     \def\2{{1\over2}} \def\4{{1\over4}}
\def\5{\bar }  \def\6{\partial } \def\7#1{{#1}\llap{/}}
\def\8#1{{\textstyle{#1}}}       \def\9#1{{\bf {#1}}}
 \def\llp{\hbox to 0pt{\hss /\hskip1.5pt}}
\def\llo{\hbox to 0.2pt{\hss /}} \def\llq{\hbox to 0pt{\hss /\hskip0.5pt}}
\def\so{\supset\hbox to 0pt{\hss $\displaystyle -$\hskip1pt}}

\def\<{\langle } \def\>{\rangle }

\let\nn=\nonumber
\def\bea{\begin{eqnarray}} \def\eea{\end{eqnarray}}
\def\beann{\begin{eqnarray*}} \def\eeann{\end{eqnarray*}}
\def\beq{\begin{equation}} \def\eeq{\end{equation}}

\date{}
\title{
{\large\rm DESY 94-045}\hfill{\large\tt ISSN 0418-9833}\\
{\large\rm March 1994}\hfill\vspace*{3.5cm}\\
Confinement in Three Dimensions\\
and the Electroweak Phase Transition}
\author{W. Buchm\"{u}ller and Z. Fodor\thanks{On leave from Institute
for Theoretical Physics, E\"otv\"os University, Budapest, Hungary}\\
{\normalsize\it Deutsches Elektronen-Synchrotron DESY, 22603 Hamburg, Germany}
\vspace*{3.5cm}\\
}
\begin{document}

\setlength{\baselineskip}{18pt}
\maketitle
\begin{abstract}
The infrared behaviour of the standard model at finite temperature is
determined by the confining phase of the SU(2)-Higgs model in three
dimensions. Due to the Landau singularity of the three-dimensional gauge
theory the perturbative treatment of the electroweak phase transition
breaks down for Higgs masses above a critical mass $m_H^c$.
Based on a renormalization group improved effective potential we find
$m_H^c\sim$ 70 GeV. The scalar self-coupling has a
Landau-type singularity also in the abelian U(1)-Higgs model,
which leads to
a breakdown of perturbation theory in the symmetric phase.
\end{abstract}
\newpage

The phenomenon of the electroweak phase transition at high temperatures
has recently attracted much attention, mostly in connection with the
problem of baryogenesis \cite{ewlinde}. Detailed studies based on the
ring-improved perturbation theory [2-6]
suggest that the phase transition is weakly first-order for Higgs boson
masses $m_H$ smaller than the  vector boson mass $m_W$. For larger Higgs masses
the perturbative treatment becomes unreliable. An alternative, well known
technique to study second-order and weakly first-order phase transitions
of condensed matter systems is
the $\epsilon$-expansion which has also been applied to the abelian and
non-abelian Higgs models \cite{ginsparg}. A recent detailed study of
the SU(2)-Higgs model based on the $\epsilon$-expansion suggests a
more strongly first-order transition than is predicted by the
ring-improved perturbation theory \cite{arnold2}.

Let us consider the SU(2)-Higgs model at finite temperature, which for
our purposes is a sufficiently accurate approximation of the full
standard model. The analysis of the phase structure is based on a
finite-temperature effective potential which, at the critical temperature
$T_c$, exhibits a barrier separating the symmetric from the broken
phase. This potential barrier, which is responsible for the first-order
transition, is essentially generated by radiative
corrections involving fields with zero Matsubara frequency \cite{ewlinde}.
This means that
the quanta of the three-dimensional theory are responsible for the
dynamics of the transition. In fact, the entire infrared behaviour of
the finite-temperature theory is determined by the corresponding
three-dimensional theory \cite{appelquist}. This suggests
a breakdown of perturbation theory near the symmetric phase due to the
infrared Landau singularity of the three-dimensional gauge theory
\cite{wetterich}.

In the following we shall study the effect of the Landau singularity
in more detail. Our analysis will be based on the $\epsilon$-expansion
and on the assumed equivalence
between the finite-temperature theory and the three-dimensional theory at
small values of the Higgs field. The comparison between a renormalization
group improved effective potential and the familiar
ring-improved one-loop
potential will then show where pertubation theory becomes
inadequate.

The SU(2)-Higgs model in $d=4-\epsilon$ dimensions is described
by the lagrangian
\beq
\cal{L}={1\over 4}W^a_{\mu\nu}W^a_{\mu\nu}+(D_{\mu}\PH)^{\dagger}
D_{\mu}\PH + V_0(\ph^2)\quad,
\eeq
where
\beq
V_0(\ph^2)={1\over 2}m^2 \ph^2 + {1\over 4}\mu^{\epsilon}\lambda \ph^4\ ,
\ \ph^2=2\PH^{\dagger}\PH\ .
\eeq
Here $W^a_{\mu\nu}$ is the ordinary field strength tensor and
$D_{\mu}=\partial_{\mu}-i\mu^{\epsilon/2} g W^a_{\mu} \tau^a/2$
is the covariant derivative; $g$ and $\lambda$ are gauge and scalar
self-coupling, respectively, and $\mu$ is the mass scale used to define
dimensionless couplings in $4-\epsilon$ dimensions. For
our discussion of the phase structure we will need the effective
potential $V(\sigma)$, $\sigma=2\PH^{\dagger}\PH$, which in $4-\epsilon$
dimensions is given by (cf. \cite{arnold2})
\beq \label{eps1}
V_1^{\epsilon}(\sigma)
= V_0(\sigma) + 3(3-\epsilon)\cal{I}_{\epsilon}({1\over4}
\mu^{\epsilon}g^2\sigma) + \cal{I}_{\epsilon}(2\mu^{\epsilon}\lambda\sigma)\ ,
\eeq
where
\beq
\cal{I}_{\epsilon}(x) = {1\over 2}
\int\frac{d^{4-\epsilon}k}{(2\pi)^{4-\epsilon}}\ln(k^2+x)\ .
\eeq
A straightforward calculation yields
\beq
\cal{I}_{\epsilon}(x) = -\frac{\mu^{-\epsilon}x^2}{64\pi^2}
\left[{2\over \epsilon}-
\ln\left(\frac{xe^{\textstyle \gamma-{3 \over 2}}}{4\pi\mu^2}\right)
+{\epsilon\over 2}\left({\pi^2\over 12}+{5\over 8}+
{1\over 2}\ln^2\left(\frac{xe^{\textstyle \gamma-{3 \over
2}}}{4\pi\mu^2}\right)\right)
+\ldots \right]\ .
\eeq
Note, that $V_1^{\epsilon}(\sigma)$ is the gauge invariant
effective potential for the composite
field $\sigma=2\PH^{\dagger}\PH$ in the broken phase, and not
the effective potential for $\PH$
in Landau gauge. At one-loop order the contributions of scalar loops
are different for the two potentials \cite{bufohe}.

We are interested in the infrared behaviour of the SU(2)-Higgs model
at finite temperature, which is determined by the effective potential
of the three-dimensional theory,
\bea \label{vthree}
V_1(\sigma) &=& V_1^{\epsilon}(\sigma)\mid_{\epsilon=1} \nn\\
&=& {1\over 2}m^2\sigma + {1\over 4}\mu\lambda\sigma^2
    -{1\over 12\pi}\left(6\left({1\over 4}\mu g^2\sigma\right)^{3/2}
        +\left(2\mu\lambda\sigma\right)^{3/2}\right)\ .
\eea
The finite-temperature potential in four
dimension in the broken phase,
$V(\Sigma,T)$, where $\Sigma=2\Phi^{\dagger} \Phi$, agrees with the
potential (6) up to
terms of order $\cal{O}(g^3 \Sigma^3/T^2)$
(cf.\ \cite{bufohe,japaka}),
\beq
V(\Sigma,T) = T V_1\left(\sigma={\Sigma\over T}\right)\ .
\eeq
Here the parameters are identified as follows,
\bea
\mu&=&T\ ,\ \lambda(\mu)=\lambda-{3\over 128 \pi}\sqrt{{6\over 5}}g^3\ ,
\ g(\mu)=g\ ,\nn \\
m^2(\mu)&=&\left({3g^2\over 16} + {\lambda\over 2}-{3\over 16\pi}
\sqrt{{5\over 6}}g^3\right)(T^2-T_b^2)\ ,
\eea
where $T_b$ is the barrier temperature.

It is expected that
the one-loop approximation becomes unreliable as $\sigma \rightarrow 0$
due to non-perturbative effects in the symmetric phase.
In $4-\epsilon$ dimensions one can improve the one-loop potential by
summing the large logarithms by means of the renormalization group.
Extrapolating this result to $\epsilon=1$, one may hope to obtain an
improvement of the one-loop result also in three dimensions. The same
extrapolation is usually performed
for the renormalization group equations of the couplings
in connection with critical phenomena.

For the one-loop potential (\ref{eps1}) one obtains
to $\cal{O}(\epsilon)$, after subtracting terms $\cal{O}(1/\epsilon)$,
\bea \label{veps1}
V_1^{\epsilon}(\sigma)&=&{1\over 2}m^2\sigma
+{1\over 4}\mu^{\epsilon}\lambda\sigma^2 \nn\\
& &+{1\over 64\pi^2}\mu^{\epsilon}\sigma^2\left[
{9\over 16}g^4\ln\left(\frac{\mu^{\epsilon}g^2\sigma e^{2/3}}{4\hat{\mu}^2}
\right)+4\lambda^2\ln\left(\frac{2\mu^{\epsilon}\lambda\sigma}
{\hat{\mu}^2}\right)\right.\nn\\
& &\left. -{\epsilon\over 2}\left(\left({3\pi^2\over 64}
+{29\over 128}\right)g^4 +\left({\pi^2\over 3}+{5\over 2}\right)
\lambda^2\right.\right.\nn\\
& &\left.\left.
+g^4 {9\over 32}\ln^2\left(\frac{\mu^{\epsilon}g^2\sigma e^{2/3}}
{4\hat{\mu}^2}\right)
+2\lambda^2\ln^2\left(\frac{2\mu^{\epsilon}\lambda\sigma}
{\hat{\mu}^2}\right)\right)\right] + \cal{O}(\epsilon^2)\ ,
\eea
where
\beq
\hat{\mu}^2 = 4\pi\mu^2e^{\textstyle {3 \over 2}-\gamma}\ .
\eeq
Contrary to the effective potential for the Higgs field $\PH$, the
mass term is not modified by the quantum corrections. Note also the
different arguments of the logarithms due to vector loops and scalar loops,
respectively.

The exact effective potential satisfies the
renormalization group equation,
\bea
\mu{d\over d\mu} V^{\epsilon}(\sigma) &=&
\left(\mu{\partial\over \partial\mu} + \mu{d\over d\mu}m^2{\partial \over
\partial m^2} + \mu{d\over d\mu}\lambda {\partial\over \partial \lambda}
+\mu{d\over d\mu}g^2 {\partial\over \partial g^2} - \gamma_{\sigma}
\sigma{\partial\over \partial \sigma}\right) V^{\epsilon}(\sigma) \nn \\
&=& 0\quad.
\eea
Inserting the one-loop potential (\ref{veps1}) into this equation
yields the one-loop
renormalization group equations for couplings and mass,
\bea
\mu{d\over d\mu}m^2 &=& -\gamma_{\sigma}m^2 \ ,\\
\mu{d\over d\mu}g^2 &=& - \epsilon g^2 + \beta_g\ ,\\
\mu{d\over d\mu}\lambda &=& -\epsilon \lambda + a g^4 + b \lambda^2
+ 2\gamma_{\sigma}\lambda\ ,\nn\\
& & a = {9\over 128\pi^2}\ ,\ b = {1\over 2\pi^2}\ .
\eea
The anomalous dimension $\gamma_{\sigma}$ of the field $\sigma$ and
the $\beta$-function of the gauge coupling cannot be obtained from
the one-loop potential. Note, that the anomalous dimensions for the
Higgs field $\PH$ and the composite field $\sigma$ are different.
For the gauge group SU(2) the $\beta$ function is
\beq
\beta_g = -{44-n\over 48\pi^2}g^4 \equiv \beta_0 g^4\ ,
\eeq
where n is the number of Higgs doublets.

Instead of $\lambda$, we shall need in the following the effective
coupling
\beq
\bar{\lambda}(\mu) = \lambda(\mu)G^2(\mu)\ ,
\eeq
where
\beq
G(\mu) = \exp\left(-\int_{\mu_1}^{\mu}{d\mu'\over \mu'}
\gamma_{\sigma}(g(\mu'))\right)
\eeq
gives the scale dependence of the field $\sigma$.
The  renormalization group equation for $\bar{\lambda}(\mu)$ is given by
the one for $\lambda(\mu)$ with $\gamma_{\sigma}=0$.

The solutions of the evolution equations for $g$ and $\bar{\lambda}$
are known exactly \cite{chen,arnold2},
\bea
g^2(\mu)&=&\frac{x^{\epsilon}g^2_1}{1 + {\beta_0\over \epsilon}
\left(x^{\epsilon}-1\right)g^2_1}\ ,\\
\bar{\lambda}(\mu)&=&y(x) g^2(\mu)\ ,
\eea
where
\bea
y(x) &=& {1\over 8 b}\left(\beta_0+c\tan\theta(x)\right)\ ,\\
\theta(x) &=& \theta_0 - {c\over 2 \beta_0}\ln\left(1 +
{\beta_0\over \epsilon}\left(x^{\epsilon}-1\right)g^2_1\right)\ ,\\
x &=& {\mu_1\over \mu}\ ,\ g_1 = g(\mu_1)\ ,\\
\theta_0 &=& \arctan\left(\frac{2by(1)-\beta_0}{c}\right)\ ,
\ c = \sqrt{(4ab-\beta_0^2)} \ .
\eea
The running couplings $g(\mu)$ and $\lambda(\mu)$ exhibit the well known
Landau pole at
\beq \label{lpole}
\mu_L = \left(-{\epsilon\over \beta_0 g_1^2}+1
\right)^{-{1/\epsilon}}\mu_1\ .
\eeq
The renormalization group equations for $g(\mu)$ and $\lambda(\mu)$
don't have a non-trivial fixed point, which is interpreted as
evidence for a first-order phase transition in the SU(2)-Higgs model
\cite{ginsparg}.

In the finite-temperature case the critical temperature $T_c$ is very
close to the barrier temperature $T_b$. Hence, the effective mass
$m$ in the three-dimensional theory is very small, and its effect on
the position of the global minimum $\sigma_c$ can be neglected.
For simplicity, we will therefore restrict our discussion to
the case $m=0$ in the
following, i.e. we shall consider the
Coleman-Weinberg model \cite{coleman} in three dimensions.

A renormalization group improved potential sums large logarithms. In
the potential (\ref{veps1}) two types of logarithms appear,
due to scalar and vector loops. Since $\lambda\ll g^2$, we neglegt
the scalar loops.
For $m=0$, the renormalization group improved potential then takes the very
simple form \footnote{A discussion of the standard model at
zero temperature is given, e.g., in ref.\ \cite{bubu}.}
\bea \label{vrgi}
V^{\epsilon}_{RGI}(\sigma)&=&{1\over 4}\bar{\mu}^{\epsilon}
\lambda(\bar{\mu})G^2(\bar{\mu})\sigma^2 \nn\\
&=&{1\over 4}\bar{\mu}^{\epsilon}\bar{\lambda}(\bar{\mu})\sigma^2\ ,
\eea
where the mass scale $\bar{\mu}$ is given by
\beq
\bar{\mu}^2 = {g^2\over 16\pi}\mu\sigma e^{\textstyle \gamma-{5 \over 6}}
\zeta\ ,
\eeq
with $\mu_1=\mu$.
Here we have included a factor $\zeta=\cal{O}(1)$,
since a change of scale does not modify the leading logarithms in
eq.\ (\ref{veps1}). Expanding the renormalization group improved
potential (\ref{vrgi}) in powers
of $g^2$, one easily verifies that the
leading logarithms in eq.\ (\ref{veps1}) have been summed.

In three dimensions the potential
$V_{RGI}\equiv V_{RGI}^{\epsilon}\mid_{\epsilon=1}$
can now be compared with the one-loop potential (\ref{vthree}).
A formal expansion in powers of $g^2$ now yields an expansion in
powers of $\sqrt{\mu/g^2\sigma}$,
\bea
V_{RGI} &=& {1\over 4}\bar{\mu}\bar{\lambda}(\bar{\mu})\sigma^2 \nn\\
&=& {1\over 4}\mu\sigma^2\left(\lambda
-{9\over 128\pi^2}g^4{\mu\over \bar{\mu}}+\ldots\right)\ ,
\eea
since $\bar{\mu}=\cal{O}(\sqrt{g^2\mu\sigma})$. The term of order
$\cal{O}(\sigma^{3/2})$ is identical with corresponding term of the
one-loop potential (\ref{vthree}) if one chooses
\beq
\zeta = {81\over 64\pi}e^{\textstyle -\gamma+{5 \over 6}}\ .
\eeq

For $m^2=0$, the one-loop potential (\ref{vthree}) has its global
minimum at $\sigma_c > 0$.
As one increases the Higgs mass, the position of the minimum moves
closer to the origin. Clearly, perturbation theory breaks down as one
approaches the Landau pole. The singularity is reached for
$\bar{\mu}(\sigma_c)=\mu_L$, where $\mu_L$ is given by eq.\ (\ref{lpole})
with $\mu_1 = \mu$. This condition for the breakdown of perturbation
theory defines a critical Higgs mass, which is given by
\beq
{{m_H^c}^2\over m_W^2} = -{3\over 8\pi\sqrt{\pi}}
e^{\textstyle {\gamma \over 2}-{5 \over 12}}{1\over \beta_0}\ .
\eeq
For the SU(2)-Higgs model one obtains $m_H^c \simeq 0.81\ m_W$ which,
with $m_W = 80$ GeV yields a critical Higgs mass of about 70 GeV.

In fig.\ (1) the one-loop potential (\ref{vthree}) and the renormalization
group improved potential (\ref{vrgi}) are compared for a Higgs mass
$m_H = 25$ GeV and the standard model gauge coupling $g^2=0.41$.
For the improved potential the minimum is shifted inwards by about 20\%,
and at  $\sqrt{\sigma_L} = 0.2\ \sqrt{\sigma_c}$ the potential
shows the Landau singularity. As fig.\ (2) demonstrates, the effect of the
Landau singularity is already very dramatic at $m_H = 45$ GeV.
Note, however, that our estimate of the location of the breakdown of
perturbation theory is uncertain at least by a factor of two.

What is the effect of the strong interactions in the symmetric phase
on the electroweak phase transition? It is conceivable that the
non-perturbative effects increase the strength of the transition,
for instance via the formation of condensates \cite{shaposhnikov,kajantie}.
Another possibility is that the main non-perturbative
effects can be parametrized by means of a ``magnetic mass'', which
would decrease the strength of the transition \cite{bufo}. At present,
only lattice simulations appear to be able to settle this issue
\cite{bunk,montvay}.

Let us finally consider the abelian Higgs model. We define the gauge
coupling via the covariant derivative $D_{\mu}\PH =
(\partial_{\mu} - i g \mu^{\epsilon/2}\ W_{\mu})\PH$ and choose for its value
and for the vector boson mass the standard model values, as in our
discussion of the SU(2)-Higgs model. All equations used for the
discussion of the non-abelian case remain the same, if we change
the parameters $a$ and $\beta_0$ in the renormalization group equations to
\beq
a = {3\over 8 \pi^2}\ ,\ \beta_0 = {1\over 24\pi^2}\ .
\eeq
Contrary to the nonabelian case, the running gauge coupling remains
finite as $\mu \rightarrow 0$. However, its value becomes rather large,
\beq
g^2(0) = \beta_0^{-1}\ .
\eeq
In the coupled system of renormalization group equations
the non-zero value of $g^2$ leads to a pole
in the effective coupling $\bar{\lambda}(\bar{\mu})$ of the
renormalization group improved potential. One finds a Landau-type
singularity at
\beq
\sqrt{\sigma} = 4\sqrt \pi \beta_0 g e^{\textstyle -{\pi\beta_0 \over c}-
{\gamma \over 2} + {5 \over 12} }\sqrt{\mu}
             \simeq 0.03\ \sqrt{\mu}\ .
\eeq
In figs.\ (3) and (4) the one-loop potential and the renormalization group
improved potential are shown for $m_H = 25$ GeV and $m_H = 45$ GeV.
Compared to the non-abelian case Landau singularity and global minimum
are more separated. However, for sufficiently large Higgs masses,
non-perturbative effects appear to be important also in the abelian case.

In summary, we have to conclude that non-perturbative effects appear
to be important in the symmetric phase of the non-abelian and also the
abelian Higgs model. In the non-abelian case this is expected, whereas
in the abelian case this may come as a surprize. In the standard model
the perturbative approach becomes inadequate for Higgs masses
above $\sim 70$ GeV.

\noindent
{\bf\large Figure captions}

\noindent
{\bf Fig.1} One-loop potential and renormalization group improved
potential for the SU(2)-Higgs model in units of $\mu=T$. $m_H=25$ GeV.

\noindent
{\bf Fig.2} One-loop potential and renormalization group improved
potential for the SU(2)-Higgs model in units of $\mu=T$. $m_H=45$ GeV.

\noindent
{\bf Fig.3} One-loop potential and renormalization group improved
potential for the U(1)-Higgs model in units of $\mu=T$. $m_H=25$ GeV.

\noindent
{\bf Fig.4} One-loop potential and renormalization group improved
potential for the U(1)-Higgs model in units of $\mu=T$. $m_H=45$ GeV.
\end{document}